\documentclass[draft]{agujournal2019}
\usepackage{url} %this package should fix any errors with URLs in refs.
\usepackage{lineno}
\usepackage[finalnew]{trackchanges} %for better track changes. finalnew option will compile document with changes incorporated.
\usepackage{soul}
\soulregister\ref7
\soulregister\cite7
\soulregister\citeA7
\linenumbers
\draftfalse

\journalname{Geophysical Research Letters}
\newcommand{\scr}[1]{_{\mbox{\protect\scriptsize #1}} }

\begin{document}
\nolinenumbers
\title{2D reconstruction of magnetotail electron diffusion region measured by MMS}

\authors{J.~M.~Schroeder\affil{1}, J.~Egedal\affil{1}, 
G.~Cozzani\affil{2}, Yu.~V.~Khotyaintsev\affil{3},
W.~Daughton\affil{4},
R.~E.~Denton\affil{5},
J.~L.~Burch\affil{6}
}

\affiliation{1}{Department of Physics, University of Wisconsin-Madison, Madison, Wisconsin 53706, USA}
\affiliation{2}{Department of Physics, University of Helsinki, Helsinki, Finland}
\affiliation{3}{Swedish Institute of Space Physics, Uppsala 75121, Sweden}
\affiliation{4}{Los Alamos National Laboratory, Los Alamos, New Mexico 87545, USA}
\affiliation{5}{Department of Physics and Astronomy, Dartmouth College, Hanover, NH,
USA}
\affiliation{6}{Southwest Research Institute, San Antonio, TX, USA}
%(repeat as many times as is necessary)

%% Corresponding Author:
% Corresponding author mailing address and e-mail address:

% (include name and email addresses of the corresponding author.  More
% than one corresponding author is allowed in this LaTeX file and for
% publication; but only one corresponding author is allowed in our
% editorial system.)

\correspondingauthor{J. M. Schroeder}{schroeder24@wisc.edu}

%% Keypoints, final entry on title page.

%  List up to three key points (at least one is required)
%  Key Points summarize the main points and conclusions of the article
%  Each must be 100 characters or less with no special characters or punctuation and must be complete sentences

% Example:
% \begin{keypoints}
% \item	List up to three key points (at least one is required)
% \item	Key Points summarize the main points and conclusions of the article
% \item	Each must be 100 characters or less with no special characters or punctuation and must be complete sentences
% \end{keypoints}

\begin{keypoints}
\item The fluctuating measurements are consistent with a 2D reconnecting geometry, permitting a detailed spacecraft trajectory to be determined.
\item The MMS data is projected onto a 2D spatial domain, revealing the fine-scale structure of the electron diffusion region (EDR).
\item The EDR includes profiles with strong gradients in fields and flows, consistent with those observed in a matched 2D kinetic simulation.
\end{keypoints}

%% ------------------------------------------------------------------------ %%
%
%  ABSTRACT and PLAIN LANGUAGE SUMMARY
%
% A good Abstract will begin with a short description of the problem
% being addressed, briefly describe the new data or analyses, then
% briefly states the main conclusion(s) and how they are supported and
% uncertainties.

% The Plain Language Summary should be written for a broad audience,
% including journalists and the science-interested public, that will not have 
% a background in your field.
%
% A Plain Language Summary is required in GRL, JGR: Planets, JGR: Biogeosciences,
% JGR: Oceans, G-Cubed, Reviews of Geophysics, and JAMES.
% see http://sharingscience.agu.org/creating-plain-language-summary/)
%
%% ------------------------------------------------------------------------ %%

%% \begin{abstract} starts the second page

\begin{abstract}

Models for collisionless magnetic reconnection in near-Earth space are distinctly characterized as 2D or 3D. In 2D kinetic models, the frozen-in law for the electron fluid is usually broken by laminar dynamics involving structures set by the electron orbit size, while in 3D models the width of the electron diffusion region is broadened by turbulent effects. We present an analysis of {\it in situ} spacecraft observations from the Earth's magnetotail of a fortuitous encounter with an active reconnection region, mapping the observations onto a 2D spatial domain. While the event likely was perturbed by low-frequency 3D dynamics, the structure of the electron diffusion region remains consistent with results from a 2D kinetic simulation. As such, the event represents a unique validation of 2D kinetic,  and laminar  reconnection models. 
\end{abstract}

\section*{Plain Language Summary}
\add[editor]{Magnetic reconnection is a fundamental process that occurs in the near-Earth space environment with implications for the safety and longevity of space-borne electronics in which magnetic field lines rearrange and release energy. To understand whether reconnection is better described as occurring in a 2D-plane without variation in the third direction versus 3D with variation in all directions, we analyze spacecraft data from the night-side of Earth's magnetic field. We conclude for the considered event that the innermost region, where the field lines reconnect, remains consistent with results from a 2D simulation.}

%Magnetic reconnection is a fundamental process that occurs often in the near-Earth space environment in which magnetic field lines rearrange and release considerable amounts of energy. One common location for reconnection is the Earth's magnetotail, the section of Earth's magnetic field on the night side of the planet. In order to understand whether reconnection can be better described as occurring in a 2D-plane with no variation in the third direction versus fully 3D with variation in all directions, we turn to spacecraft data. We present an analysis of spacecraft observations from the Earth's magnetotail of a fortuitous encounter with an active reconnection region, mapping the observations onto a 2D spatial domain. While the event likely was affected by 3D dynamics, the structure of the small-scale innermost region, where the field lines reconnect, remains consistent with results from a 2D simulation. As such, the event represents a unique validation of 2D kinetic reconnection models.

\section{\label{sec:intro}Introduction }

Magnetic reconnection \cite{dungey:1953} is a fundamental physical process in plasmas in which magnetic field lines rearrange their topology and convert magnetic energy into particle thermal and kinetic energies \cite{zweibel2009}. Despite a large body of research on this topic, the precise physics of what occurs at the smallest scales in these reconnection events is yet to be fully understood. One outstanding question is whether structures of the innermost reconnection region where the electron fluid decouples from the magnetic field, called the electron diffusion region (EDR), are best described by laminar, 2D kinetic models \cite{vasyliunas:1975,pritchett:2001} or by 3D models that predict instabilities that broaden features of the reconnection site \cite{papadopoulos:1977,huba:1977,hoshino:1991}. To address this issue, NASA recently launched the Magnetospheric Multiscale Mission (MMS), which orbits about Earth’s magnetosphere and is designed specifically to probe the small-scale structure of magnetic reconnection in situ \cite{burch:2016b}. \remove[editor]{Initial EDR encounters measured by MMS indicated a laminar structure dominated by electron dynamics  \cite{burch_edr_2016,torbert_edr_2018} }. We here revisit one out of several electron diffusion regions observed by MMS in the Earth's magnetotail, with the aim to compare the recorded dynamics to numerical results obtained with a 2D kinetic simulation model.

% { \if \cite{burch_edr_2016,torbert_edr_2018} \fi }

\section{Summary of MMS event}

On August 10, 2017 MMS observed a reconnection event in Earth's magnetotail. During this event the spacecraft flew fortuitously along a topological boundary between plasma inflow and outflow, also known as a magnetic separatrix, and straight through the center of a reconnection EDR. The event has been the subject of previous investigations \cite{zhou2019,hasegawa2022,denton2020,cozzani2021}. In \citeA{cozzani2021}, the frequency of strong fluctuations in fields and flows are found to be consistent with kinking of the current sheet in the direction normal to the reconnection plane at the lower-hybrid frequency. In addition, the signals could also be influenced by  the reconnection exhaust separatrices being rippled by small magnetic islands \cite{denton2020}. While 3D geometry and time-dependent dynamics are suggested to describe the observation, whether the nature of the event can be accounted for by a 2D laminar framework is still not fully determined. Considering the same event, we here explore in greater detail the evidence for 2D or 3D reconnection dynamics. We seek to address the extent to which the observed signals are consistent with a 2D  reconnection geometry that is being ``pushed around'' as a rigid body by Alfv\'enic perturbations external to the EDR. The stronger fluctuations observed in the vicinity of the EDR could then be interpreted as the result of the spacecraft zigzagging through the reference frame of a static 2D geometry with strong spatial gradients, rather than caused by a localized 3D instability.

In Fig.~\ref{fig:rawdata}\change[editor]{(a-d)}{(a-e)} several key measurements by MMS are shown in the time interval of the reconnection encounter. The observation of a reversal in the  $B_L$ magnetic field component (seen near 12:18:33 UTC in panel a) bracketed by strong electron temperature anisotropy (panel b) indicate an EDR encounter where the MMS path sampled the EDR as well as the boundaries of the two opposing inflow regions, since reconnection inflows are typically characterized by enhanced parallel electron temperature. Additionally, the normal electric field and electron outflow velocity profiles include significant fluctuations shortly after the $B_L$ reversal. The times within the grey dashed bars in panels \change[editor]{a-d}{a-e} indicate the region of interest and are chosen to encompass the features described above. We will call this interval the reconnection region, as MMS samples both the EDR and along the magnetic separatrix. The reconstructed signals in Fig.~\ref{fig:rawdata}(f-j) will be discussed in further detail at the end of the manuscript, but to the extent that they agree with the corresponding signals in Fig.~\ref{fig:rawdata}(a-e), they show that a 2D reconnection geometry is consistent with the observation.

\begin{figure}
	\centering
\includegraphics[trim=30mm 5mm 25mm 5mm, width=0.85\textwidth]{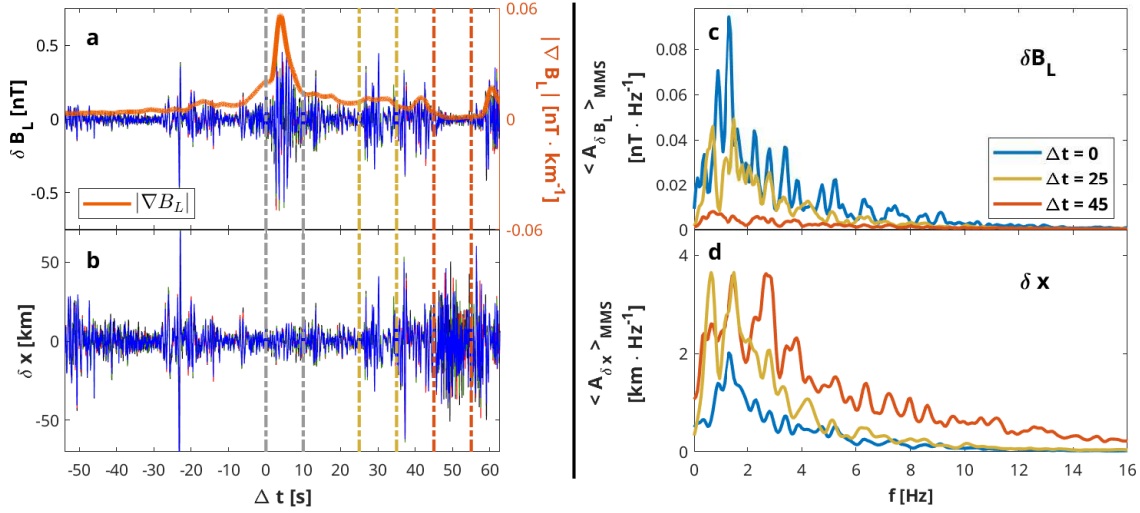}
\caption{\label{fig:blflucts_psd} (a) Fluctuations in the magnetic field in the {\it L}-direction, $\delta B_L$, for each spacecraft are shown. A peak in fluctuations can be seen near $\Delta t$=5. We compute the high-pass filtered signal by subtracting the Gaussian-smoothed mean signal from the measurement, using a Gaussian with a width of 1s. (b) The plasma excursion, $\delta {\bf x} = \delta B_L/|\nabla B_L|$. (c) Amplitude spectra of fluctuations in $B_L$ over 10 second time intervals averaged across all four spacecraft, $\langle A_{\delta  B_L} \rangle\scr{MMS}$. Shown are three spectra, one starting at the current sheet crossing $\Delta$t=0s and the other two at later times of $\Delta$t=25s and $\Delta$t=45s. (d) $\langle A_{\delta {\bf x}} \rangle\scr{MMS}$ for the same three time intervals considered in (c).}
\end{figure}

To examine the characteristics of the fluctuations present in the EDR, in Fig.~\ref{fig:blflucts_psd}(a) the high-pass filtered $B_L$ signal is shown for a wider time interval, where $\Delta t$=0\remove[editor]{,} corresponds to 12:18:30 UTC. The $B_L$ fluctuations\add[editor]{, denoted $\delta B_L$,} exhibit a notable increase in amplitude in the reconnection region. We first seek to characterize the gradient length scale typical for various quantities across the reconnection encounter. In turn, this will allow us to compute the level of displacement $\delta {\bf x}$ required by a hypothetical 2D rigid geometry to account for the observed fluctuation levels.

\remove[editor]{The spatial gradients in a given MMS field can be estimated by root-mean-square of differences in the signal measured by each spacecraft from the mean value across all four spacecraft. Let $\Delta S_n \equiv  S_n -  \overline{S}({\bf\vec{x}} )$ denote the difference between the  MMS signal for spacecraft $n$ and the  mean signal $\overline{S}(\bf{\vec{x}})$, representative of the field at the  centroid $\bf{ \vec{x}}$ of the MMS tetrahedral formation. The signal from a single spacecraft is then approximately  $\overline{S}({\bf\vec{x}}) + \nabla S({\bf\vec{x}}) \cdot \vec{d \ell}_n$, where $\vec{d \ell}_n$ is the location of MMS-n relative to the centroid. 
Taking the root-mean-square of $\Delta S_n$, it then follows that
\protect\begin{equation}
\label{eq:grad}
     |\nabla S({\bf\vec{x}})| \sim \Delta S\scr{rms}\ /\ |\overline{d \ell}|
\end{equation} } \remove[editor]{where $|\overline{d \ell}|$ is the average distance of each spacecraft from the tetrahedral centroid.}

\change[editor]{ Computed using Eq.~\ref{eq:grad}, in Fig.~\ref{fig:blflucts_psd}(a) the orange curve represents the magnitude of the spatial gradient in $B_L$.}{In Fig.~\ref{fig:blflucts_psd}(a) the orange curve represents the magnitude of the spatial gradient in $B_L$ calculated to linear order using low-pass filtered data \cite{MutlispacecraftAnalysis}. } Normalizing the fluctuations by this gradient scale yields a quantity denoted as $\delta {\bf x} = \delta B_L/|\nabla B_L|$ in Fig.~\ref{fig:blflucts_psd}(b), which represents the plasma excursion projected along $\nabla B_{L}$.
Significant to the present analysis, we observe that this displacement measure, $\delta {\bf x}$, has approximately constant amplitude inside and outside of the reconnection region. 
To further address if the displacement amplitude is uniform or peaked at the reconnection region, we examine the amplitude Fourier spectrum for the fluctuations in Figs.~\ref{fig:blflucts_psd}(a,b) at different time intervals. In panel (c) we show $\langle A_{\delta  B_L} \rangle\scr{MMS}$, the mean amplitude of the $B_L$ fluctuations for all four spacecraft, over three 10 second time intervals. The first of these, beginning at $\Delta t=0$s, covers the crossing of the x-line and the most significant fluctuation amplitude in $B_{L}$ (between the grey dashed lines in panels (a,b), approximately matching the grey reconnection interval in Fig.~\ref{fig:rawdata}). The two other time intervals are chosen to start at $\Delta t=25$s and $\Delta t=45$s, which we estimate to be  about 5$d_i$ and 9$d_i$ from the x-line, respectively, based on extrapolating the velocity of the spacecraft trajectory discussed in the next section ($d_i=c/\omega_{pi}$ is the ion inertial length). In panel (d) we compute $\langle A_{\delta {\bf x}} \rangle\scr{MMS}$ over the same time intervals. All spectra exhibit a noisy profile of spectral amplitude peaking around 1-2 Hz. While the $B_{L}$ fluctuations have considerably stronger power during the central current sheet crossing, the excursion $\delta {\bf x}$ appears larger outside the EDR. It should be noted that the second time interval occurs at times with significant $B_N$ variation and the third over a quieter period. Time intervals over different field structures display different spectral features from those chosen \change[editor]{here; however, that the normalized $\delta B_L$ fluctuation amplitude becomes comparable to the fluctuations of other time intervals remains true.}{here, but importantly the $\delta {\bf x}$ fluctuation amplitude at the current sheet crossing does not dominate over other intervals in the same way as $\delta B_L$. We additionally repeated this analysis with other components of the magnetic field and found that gradients in those quantities are much smaller and lead to unrealistically high values of $\delta {\bf x}$; however, other quantities such as fluctuations in electron flow speed yield similar results.}

The observations based on Fig.~\ref{fig:blflucts_psd}  provide evidence that the fluctuations observed within the EDR are influenced by large-scale fluctuations of the magnetotail current sheet not generated  by the reconnection event itself. From Fig.~\ref{fig:blflucts_psd}(d), using the peak frequency $f\simeq1.5$Hz and the mean amplitude $\langle A_{\delta {\bf x}} \rangle\scr{MMS}\simeq2.75$km/Hz, we estimate a typical perturbation speed as $v_{fluct}= 2\pi f \langle A_{\delta {\bf x}} \rangle\scr{MMS} \simeq 25$km/s. This speed is small compared to the upstream Alfv\'en speed, $V_{A\infty}\simeq 400$km/s, providing additional evidence that the fluctuations observed within the EDR are imposed by modest Alfv\'enic perturbations existing throughout the reconnecting current sheet. These perturbations likely include kinking of the current sheet at scales larger than the 2D plane as discussed in \citeA{cozzani2021} and possibly other types of Alfv\'enic waves. The present small level of activity is also in contrast to the event in \citeA{li_egedal_2021}, in which fluctuations of $\delta {\bf E}$ near the EDR were elevated by orders of magnitudes above the background level of upper hybrid waves. In addition to externally imposed fluctuations, below we find that features in the $B_N$ measurement indicate that magnetic islands as identified in \citeA{denton2020} may also be a source for ripples in the separator.

\section{Comparison to kinetic simulation}

 We now explore the extent to which the observed signals are compatible with results from a 2D kinetic simulation. 
 We apply VPIC particle-in-cell code \cite{bowers:2009}, which is used extensively to study magnetic reconnection. As in \citeA{cozzani2021}, the numerical run initialized with a normalized upstream plasma beta $\beta_{e,\infty}=0.09$, temperature ratios $T_{i,\infty}/T_{e,\infty}=5$, and a guide-field $B_g/B_0=0.1$. However, compared to  \citeA{cozzani2021} we use an earlier time slice, which has a 30\% lower reconnection rate and reduced separator angles yielding better agreement with the observations.  The simulation uses the full  proton to electron  mass ratio, $m_i/m_e=1836$,  which allows for a direct quantitative mapping between VPIC and MMS units \cite{egedal2019}. As discussed in \citeA{le:2013}, the guide magnetic field $0.05< B_g/B_0 <0.15$ renders this a Regime II event  with no extended electron jets expected to emanate beyond a localized EDR.

\begin{figure}
\centering
\includegraphics[trim=20mm 5mm 15mm 5mm, width=0.7\textwidth]{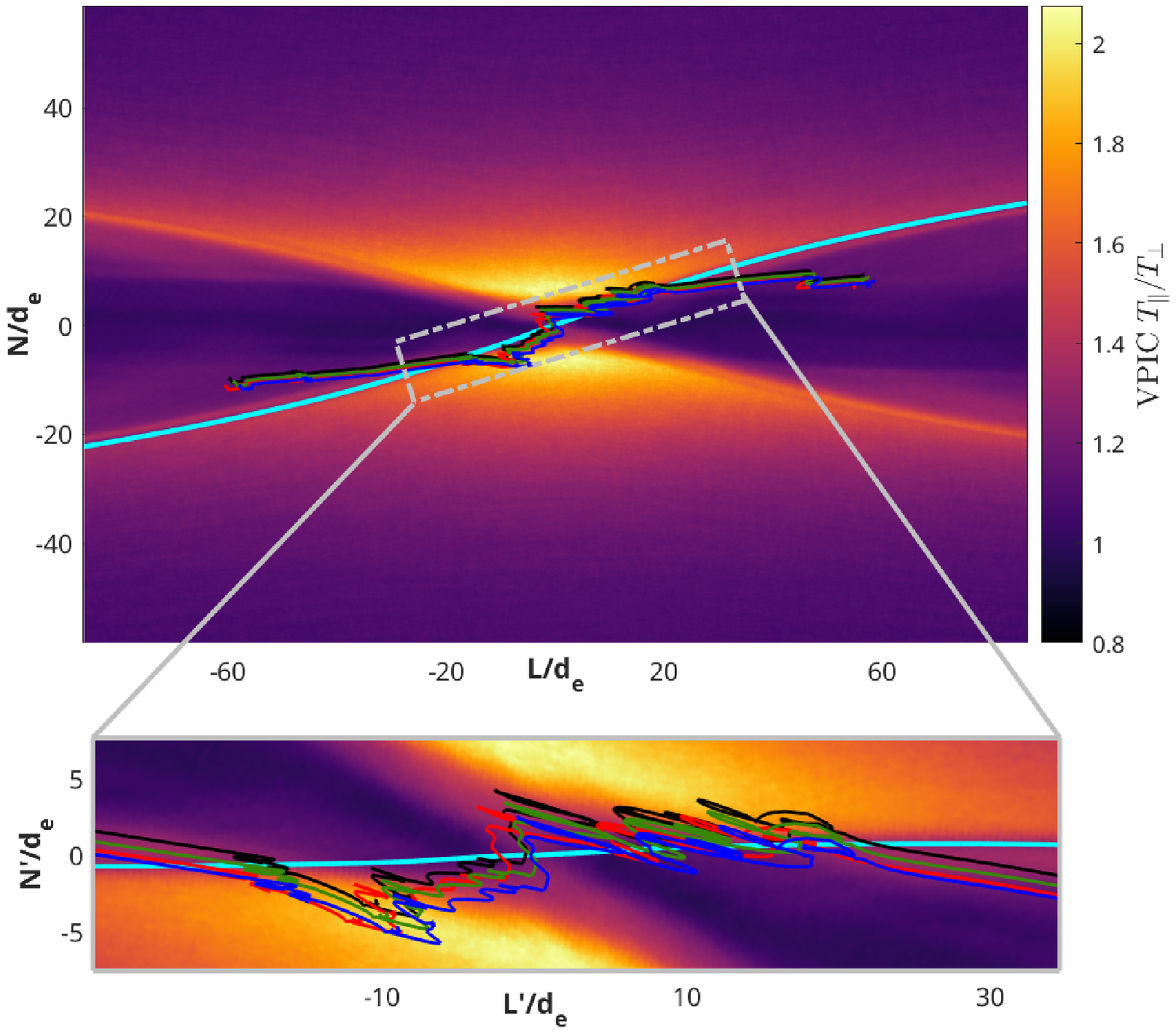}
\caption{\label{fig:mmspath} Upper panel: The spacecraft paths are plotted over VPIC simulation data of the ratio of electron temperatures parallel and perpendicular to the magnetic field. One of the two magnetic separatrices, plotted as a cyan curve, is used as the basis of a coordinate transform in 2D MMS plots shown in Fig.~\ref{fig:2dfields}. Lower panel: Zoomed in view of the domain and coordinates used for 2D reconstruction.}
\end{figure}

	 In Fig.~\ref{fig:mmspath} the inferred spacecraft paths are plotted over VPIC data for the electron temperature anisotropy, characterized by the ratio of electron temperature parallel and perpendicular to magnetic field lines. 
	 The paths are found through a least-squares optimization procedure, similar to that laid out in \citeA{egedal2019}. 
	 However, for the present analysis, the MMS1 $B_L$ signal directly determines the simulation $B_L$ contour along which the spacecraft position is optimized. Further details of this procedure are discussed in Section 1 of the Supporting Information accompanying this text.

	\begin{figure}
	\centering
\includegraphics[trim=25mm 5mm 15mm 5mm, width=0.85\textwidth]{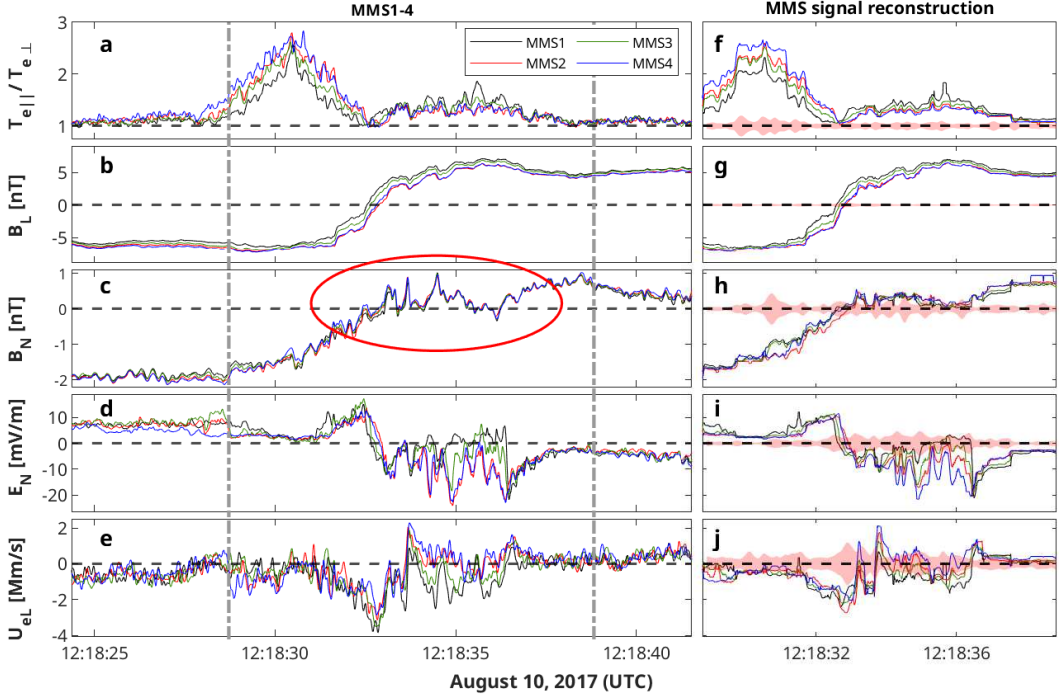}
\caption{\label{fig:rawdata} Time series data from the magnetotail reconnection event measured on August 10, 2017. Shown from top to bottom are (a) temperature anisotropy, (b/c) magnetic field in the {\it L/N}-direction, (d) electric field in the {\it N}-direction, and (e) electron velocity in the {\it L}-direction. We use the previously determined event basis of \citeA{denton2020}, where unit vectors in geocentric solar ecliptic (GSE) coordinates are [{\bf L};{\bf M};{\bf N}] = [0.9872,   -0.1305,   -0.0915;
    0.1580,    0.8782,    0.4515;
    0.0214,   -0.4601,    0.8876], aligning the separatrix in the MMS event with the corresponding separatrix in the VPIC simulation.  
    (f-j) show the result of interpolation through 2D maps of Fig.~\ref{fig:2dfields} along the spacecraft paths shown in Fig.~\ref{fig:mmspath}. Shaded in pink in is the reconstruction error, taken as the root-mean-square of the differences between the measured signal and the reconstructed signal for MMS1-4.
}
\end{figure}

\begin{figure}
\centering
\includegraphics[trim = 18mm 5mm 18mm 10mm,width=\columnwidth]{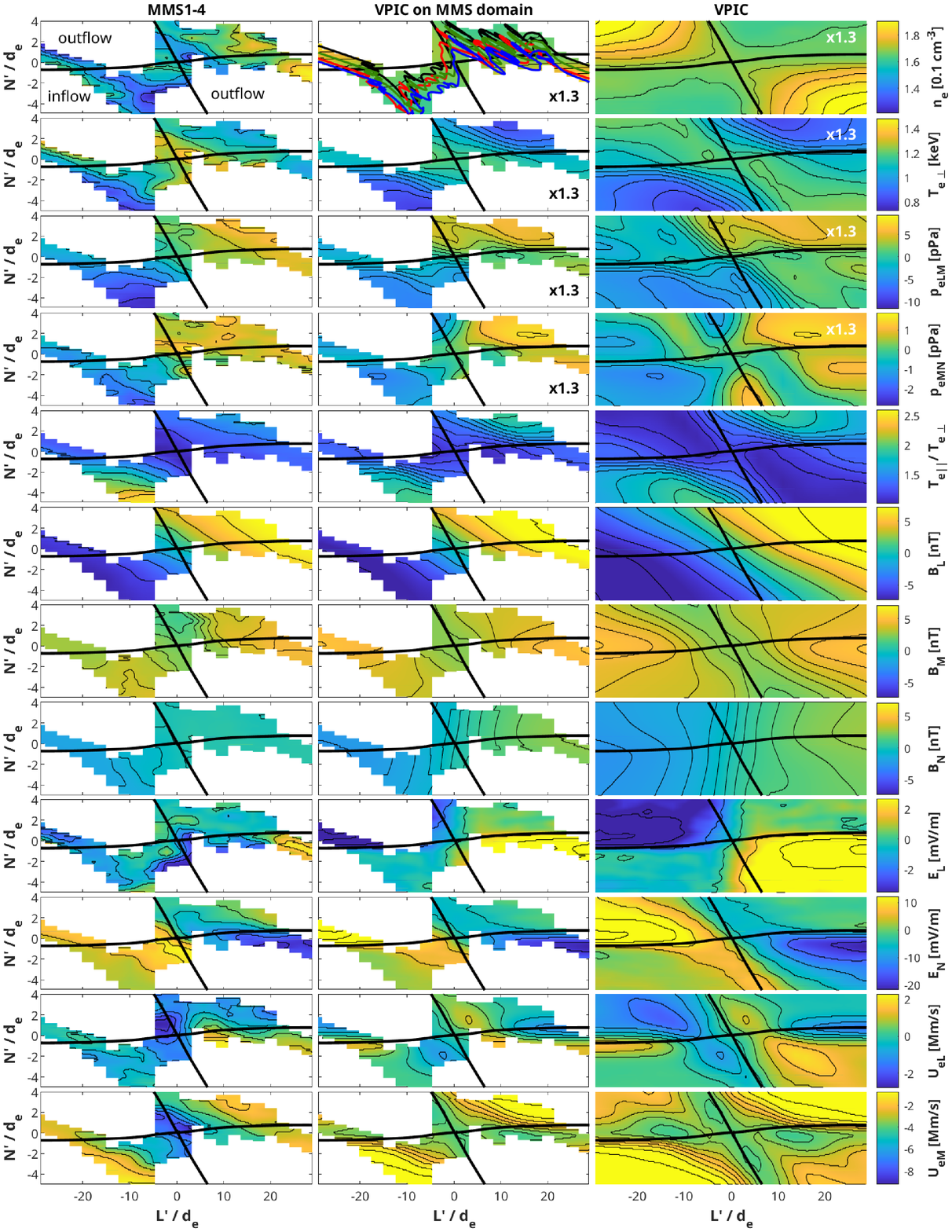}
\caption{\label{fig:2dfields} The left column shows fields measured with MMS constructed into a 2D map based on the spacecraft trajectory. The map is given in rotated coordinates ($L', N')/d_e$, where $L'$ approximately represents the distance along and $N'$ the distance away from the separatrix highlighted in cyan in Fig.~\ref{fig:mmspath} (shown here as a horizontal black curve). The middle column shows VPIC data of the same region of the reconnection geometry for comparison, and the right column shows VPIC data over the entire domain plotted. The top panel of the VPIC data in the middle column is overlaid with the spacecraft paths in the rotated coordinates. The first four fields of VPIC data are multiplied by a factor of 1.3 for better visual comparison to the MMS data.}
\end{figure}

	With the paths of the spacecraft determined, the raw time series data from all four can be spatially binned and averaged to construct 2D maps of the event in the {\it LN}-plane. Technical details of this procedure are laid out in Section 2 of the Supporting Information document. The result is shown in the leftmost column of Fig.~\ref{fig:2dfields}, where MMS reconstructions for twelve different spacecraft measurement fields are shown. The corresponding VPIC simulation data are shown \change[editor]{the the}{in the} middle and right columns. The middle and right columns contain the same VPIC data values, but the middle column is restricted to the same domain as the MMS data for better visual comparison. The cyan contour of Fig.~\ref{fig:mmspath} becomes the nearly horizontal black line in the reconstructed maps. The other simulation separatrix, not shown in Fig.~\ref{fig:mmspath}, is included to highlight inflow and outflow regions distinctly. These contours represent the exact simulation separatrices and are laid over the MMS panels for reference, not to indicate the physical separatrices for the observed event. Lastly, the spacecraft paths in the rotated coordinates are superimposed on the first VPIC panel in the second column.

	\begin{figure}
\centering
\includegraphics[trim = 10mm 10mm 15mm 15mm,width=\columnwidth]{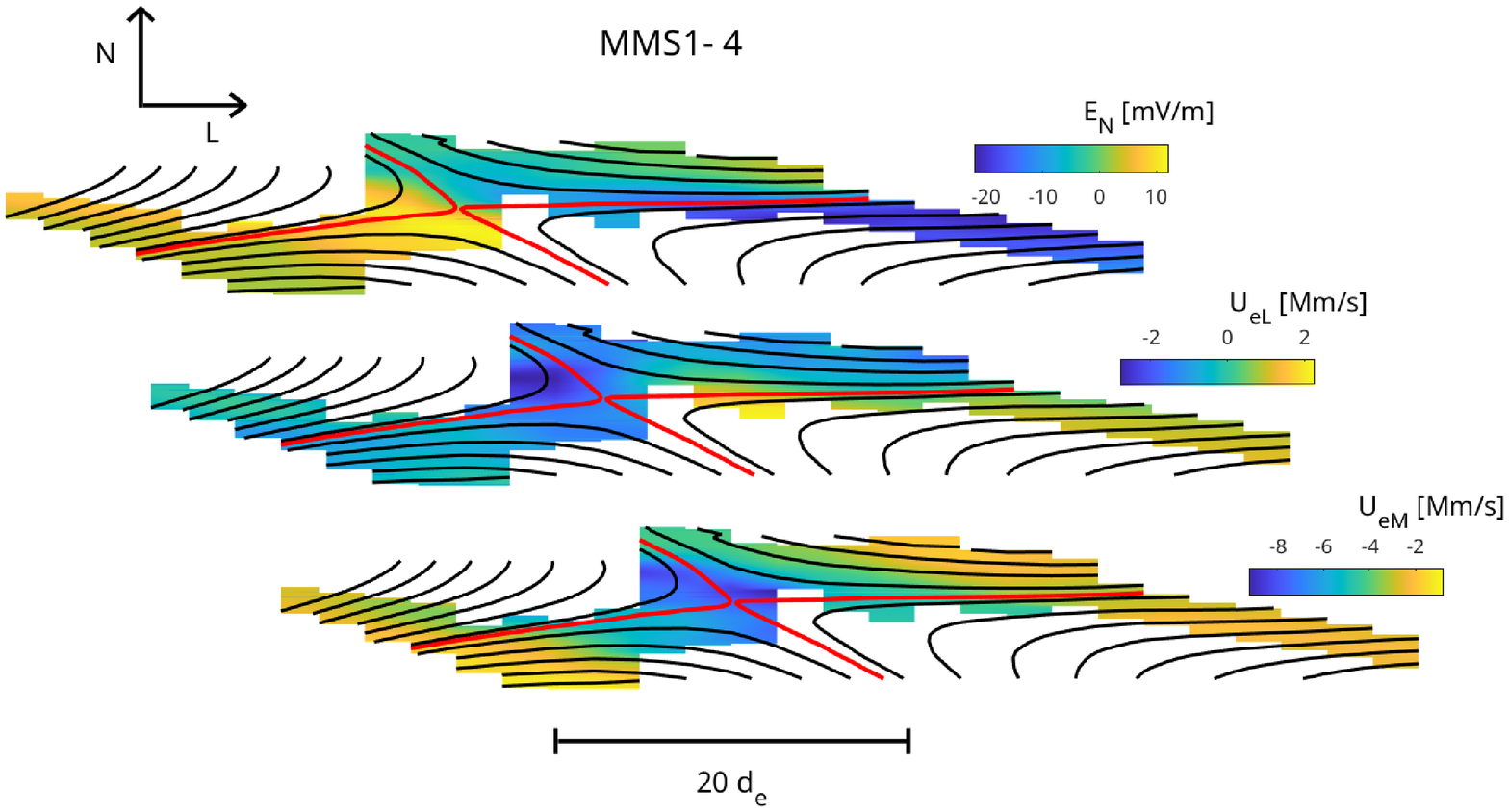}
\caption{\label{fig:flux_contours} MMS maps from Fig.~\ref{fig:2dfields} in {\it L'N'}-coordinates. 
Overlaid are magnetic field lines in the {\it LN}-plane calculated from  $A_M$ contours, where $A_M$ is obtained directly from MMS 2D maps of $B_L$ and $B_N$. Contours that fall outside of underlying color plot come from interpolation of magnetic field maps along the {\it L}'-direction.  
}
\end{figure}

Visual comparison of the reconstructed maps to the simulation data shows striking similarity in many key features in each quantity. Sharp gradients across the separator around $L'/d_e \sim$ 10-25 are seen in multiple fields including $E_N$ where a variation along {\it N}' of $\sim25$mV/m over $\sim 2 d_e$ matches excellently with the simulation data. As described above, at each time-point, the optimization for the four spacecraft locations only includes a single degree of freedom. Therefore, the good match between multiple MMS maps and VPIC profiles is direct evidence that the EDR has a rigid 2D structure, without strong 3D effects or fast temporal variations. For example, had the event included significant 3D dynamics that broaden separatrix layers, sharp gradients in multiple quantities (including the elements of ${\bf E}$ and ${\bf u_{e}}$) would be smoothed out by the overlapping {\it LN}-projected spacecraft paths. Good quantitative agreement is also observed between the MMS and VPIC quantities, but to fit a common color map, the VPIC profiles of $n_e$, $T_{e\perp}$, $P_{eLM}$, and $P_{eMN}$ in Fig.~\ref{fig:2dfields} are each multiplied by a factor of 1.3. This suggests that the  value of $\beta_{e\infty}$ applied in the simulation could still be optimized further. However, this simulation parameter is not expected to influence the accuracy of the presented 2D MMS data profiles.

\change[editor]{To further illustrate the how between the 2D spacecraft data presented in Fig.~\ref{fig:2dfields} well represent the MMS observations, time series data directly comparable to MMS signals can be constructed by interpolating the MMS maps along each respective spacecraft path.}{To further illustrate how the 2D reconstructed maps well represent the observation, we construct time series data directly comparable to the MMS signals by interpolating the 2D maps along each respective spacecraft path.} This is presented in Fig.~\ref{fig:rawdata}\change[editor]{(e-h)}{(f-j)}, where the measured MMS signals of four key fields are in the left column while the interpolated signals are shown in the right column. Good agreement is seen for each field, confirming the accuracy of the reconstructed 2D maps of the event. The pink shaded regions in the second column are error estimates for the reconstruction, the root-mean-square of the difference between MMS measurement and the reconstructed signal across all four spacecraft. The error estimate for $B_L$ is particularly small because that quantity is fixed to the MMS1 measurement in the trajectory optimization (see Supporting Information, Section 1). In other fields, however, the error remains bounded to a small value for all times and thus indicates the merit of the 2D reconstruction method. 
	
Having shown that a 2D geometry well suits the event considered, we now consider the in-plane magnetic field profile and separatrix locations for the spacecraft observation. The in-plane magnetic field fulfills ${\bf B}_{LN} = \nabla \times A_M {\bf e}_M$, where $A_M$ is the $M$-component of the magnetic vector potential, so it follows that $A_M(L,N) = \int_{L_0,N_0}^{L,N}  {\bf B}_{LN}  \times \vec{d \ell}_{LN}$ \cite{kesich:2008}. Here the integral can  be carried out along an arbitrary path from an arbitrary starting point $(L_0,N_0)$. In Fig.~\ref{fig:2dfields} the MMS profiles of $B_L$ and $B_N$ are relatively smooth and we can interpolate where possible in the $L'$ direction to obtain an enhanced area of coverage. $A_M(L,N)$ is then obtained by first integrating from $L'/d_e=-30$ to $L'/d_e=30$ along the center of the domain where ${\bf B}_{LN}$ data is available, characterizing $A_M(L',N')$ along this center-line. Any point in the $(L',N')$-plane is then reached by integrating in $N'$-direction from the center-line. 

The result is shown in Fig.~\ref{fig:flux_contours}, where black lines represent contours of constant $A_M$ and the red lines represent the topological separatricies. To the best of our knowledge, this is the first map of the field lines in a region of an EDR obtained directly by MMS data without relying on any extrapolation techniques. The present magnetic map is based on all the data of the EDR encounter and covers a spatial domain of about 60$d_e$ along the separatrix. 

We observe how the separatricies mark the locations of the strongest gradients in $E_N$, $U_{eL}$ and $U_{eM}$, where the gradient length scale for these quantities is at or below $1d_e$. Such fine-scale structures are not expected in scenarios including strong instabilities and/or 3D dynamics, \change[editor]{and their presence thus provides additional validation of the 2D structure assumed in the the analysis.}{which have been shown to broaden reconnection current layers by providing an additional source of electron diffusion \cite{graham_LHDI}. Their presence thus provides additional validation of the 2D structure assumed in the the analysis.} 

However, the spatial resolution is limited by the bin-size $(\simeq 2d_e)$ applied in the $L'$-direction, and this resolution does not allow us to identify magnetic islands at this scale. This can also be seen in Fig. \ref{fig:rawdata}(c/h), where the MMS measurements of $B_N$ display oscillations including sign reversals around times 12:18:32-12:18:37 (circled in red). These oscillations in the {\it N}-direction  are likely due to island structures previously reported \cite{hasegawa2022,denton2020}. Meanwhile, the $B_N$ oscillations are not as prominent in the reconstructed signals averaged out by the finite resolution of our analysis. The oscillations not captured in the reconstruction are observed to have a magnitude of about $\Delta B_N\simeq 0.5$ nT, which should be compared to $B_L\simeq 5$nT. In the $L'$-direction the ``wave-length" of fluctuations about the separatrix is about $\lambda\simeq 5 d_e$, such that the $B_N$ fluctuations correspond to a $(0.5/5) 5 d_e = 0.5 d_e$ ripple in the separatrix, not captured by the reconstruction. Thus, this effect alone is unlikely to account for the inferred magnitude of oscillations $(\simeq 4 d_e)$ in the spacecraft trajectories. 

\section{Conclusions}

In summary, we apply particle-in-cell simulation to analyze the Aug. 10, 2017 EDR event. Direct quantitative comparison provides optimized spacecraft trajectories that give spatial information in the 2D reconnection plane. Reconstruction of measured quantities in this plane provide a unique and detailed picture of the electron diffusion region with sub-electron scale resolution. The 2D reconstructions fit the underlying simulation with good accuracy suggesting that the present event has an inherently 2D-nature in agreement with results from 2D kinetic simulations. \add[editor]{Our results are also consistent with results of \citeA{zhou2019}, which conclude the event is mostly laminar with weak wave activity.} The fluctuations in signals observed are well accounted for by the short gradient length-scale $(< 1d_e)$ of the EDR, which we characterize as a rigid body being jostled by a bath of large-scale, low-amplitude fluctuations. The spatial displacements associated with these fluctuations are consistent with Alfv\'enic fluctuations observed external to the EDR, and while magnetic islands likely are present within the EDR, the data suggest that external fluctuations are responsible for the oscillatory motion of the spacecraft through the rigid 2D reconnection geometry. The reconstructed flux-function as well as the agreement of the 2D reconnection structure with those of the simulation provide a unique confirmation of laminar 2D kinetic models for reconnection.

\acknowledgments
\remove[editor]{The Magnetospheric Multiscale Mission (MMS) data used in this paper is publicly available from the CU-Boulder Laboratory for Atmospheric and Space Physics MMS Data Center webpage (https://lasp.colorado.edu/mms/sdc/public/).
Using the initial conditions specified in the text,
	the numerical data can be reproduced with the open source VPIC code available at https://github.com/lanl/vpic, as well as at \\
	\noindent
	https://zenodo.org/record/4041845\#.X2kA1x17kuY (doi - 10.5281/zenodo.4041845).}The work was funded by an 
	H. I. Romnes Faculty Fellowship by the UW-Madison Office of the Vice    Chancellor for Research and Graduate Education.
	
\section*{Open Research}	
\add[editor]{The Magnetospheric Multiscale Mission (MMS) data used in this paper is publicly available from the CU-Boulder Laboratory for Atmospheric and Space Physics MMS Data Center webpage (https://lasp.colorado.edu/mms/sdc/public/).
Using the initial conditions specified in the text,
	the numerical data can be reproduced with the open source VPIC code available at https://github.com/lanl/vpic, as well as at \\
	\noindent
	https://zenodo.org/record/4041845\#.X2kA1x17kuY (doi - 10.5281/zenodo.4041845).}

\bibliography{subelc_paper}

\end{document}